Limitation of atmospheric composition

by combustion-explosion in exoplanetary atmospheres

Grenfell, J. L.[1#], Gebauer, S.[1], Godolt, M.[2], Stracke, B.[1], Lehmann, R.[3], and Rauer, H.[1,2]

[1] Abteilung Extrasolare Planeten und Atmosphären (EPA),
Institut für Planetenforschung (IP),
Deutsches Zentrum für Luft- und Raumfahrt (DLR), Rutherford Str. 2, 12489 Berlin, Germany

[2] Zentrum für Astronomie und Astrophysik (ZAA),
Technische Universität Berlin (TUB), Hardenbergstr. 36, 10623 Berlin, Germany

[3] Alfred-Wegener-Institut,
Helmholtz-Zentrum für Polar- und Meeresforschung,
Telegrafenberg A43, 14473 Potsdam, Germany

[#]Corresponding author, contact details:

Email: lee.grenfell@dlr.de

Tel: +49 30 67055 7934

FAX: +49 30 67055 384

Abstract: *This work presents theoretical studies which combine aspects of combustion and explosion theory with exoplanetary atmospheric science. Super-Earths could possess a large amount of molecular hydrogen depending on disk, planetary and stellar properties. Super-Earths orbiting pre-main sequence-M-dwarf stars have been suggested to possess large amounts of $O_2(g)$ produced abiotically via water photolysis followed by hydrogen escape . If these two constituents were present simultaneously, such large amounts of $H_2(g)$ and $O_2(g)$ can react via photochemistry to form up to ~10 Earth oceans. In cases where photochemical removal is slow so that $O_2(g)$ can indeed build-up abiotically, the atmosphere could reach the combustion-explosion limit. Then, $H_2(g)$ and $O_2(g)$ react extremely quickly to form water together with modest amounts of hydrogen peroxide. These processes set constraints for $H_2(g)$, $O_2(g)$ atmospheric compositions in Super-Earth atmospheres. Our initial study of the gas-phase oxidation pathways for modest conditions (Earth's insolation and ~ a tenth of a percent of $H_2(g)$)) suggests that $H_2(g)$ is oxidized by $O_2(g)$ into $H_2O(g)$ mostly via HOx and mixed HOx-NOx catalyzed cycles. Regarding other atmospheric species-pairs we find that $(CO-O_2)$ could attain explosive-combustive levels on mini gas planets for mid-range C/O in the equilibrium chemistry regime (p>~1bar). Regarding $(CH_4-O_2)$, a small number of modeled rocky planets assuming Earth-like atmospheres orbiting cooler stars could have compositions at or near the explosive-combustive level although more work is required to investigate this issue.*

**1. Introduction and Motivation**

The atmospheres of Super-Earths (SEs) orbiting M-dwarf stars could build-up large amounts of abiotically-produced oxygen ($O_2$) via photolyis of water followed by escape of atomic hydrogen during the pre-main sequence phase (Luger and Barnes, 2015). Planetary formation studies (e.g. Chian and Laughlin, 2013) suggest that SEs could retain large amounts of molecular hydrogen ($H_2$) from the protoplanetary disk. The current work proposes that Combustion-Explosion (CE) reactions (e.g. Cohen, 1992) could limit the composition of exoplanetary atmospheres depending on pressure (p) and temperature (T) if a fuel gas (such as molecular hydrogen, $H_2$) is present together with an oxidant gas (such as molecular oxygen, $O_2$) initiated by lightning or cosmic rays. Whether SEs could reach the CE limit depends on the planetary evolution, specifically the timescales for photochemical oxidation, escape etc. which impact the abundance of potential combustants such as $H_2(g)$ and $O_2(g)$.

Water delivery and migration of SEs orbiting in the HZ is rather contested (Raymond et al., 2007, Ogihara and Ida, 2009). For SEs in the Habitable Zone (HZ) of M-dwarf stars explosion-combustion could represent an important mechanism by reaction of $H_2(g)$ and $O_2(g)$ for generating oceans. These would likely however contain some hydrogen peroxide ($H_2O_2$) (see CE mechanism in section 4.1) which is unsuitable for life as we know it. Also, gaseous mixtures containing ($CO-CH_4-O_2-N_2$) could possibly explode or combust on some Mini Gas Planets (MGPs) (see section 7) and on some Earth-like worlds orbiting cooler M-stars if methane builds up to above ~(2-3)% by volume (see section 8).

Section 2 considers chemical disequilibrium, combustion and life and evidence for combustion of $O_2(g)$ in early Earth's atmosphere. Section 3 discusses the initiation of CE by lightning and cosmic rays in (exo)planetary atmospheres. Section 4 reviews CE processes for ($H_2-O_2$) atmospheres. Sections 5 and 6 present the run scenario and results respectively from a model study of [$H_2-O_2$] atmospheres. Sections 7, 8 and 9 discuss potential CE for ($CO-O_2$), (Hydrocarbon-$O_2$) and ($NH_3$-containing) gas mixtures respectively. Section 10 briefly discusses explosions related to atmospheric dust suspensions. Section 11 presents a brief discussion and conclusions.

**2. Chemical disequilibrium, (atmospheric) combustion and life**

Is it possible that only planets with life could produce sufficient chemical disequilibrium to generate combustion, as occurred on the early Earth? Simoncini et al. (2013) calculated that 0.67 terawatts (TW) are required to maintain the ($CH_4(g)-O_2(g)$) redox disequilibrium in modern Earth's atmosphere with 0.24 TW associated with abiotic geological processes. They noted that this value is negligible compared with the modern Earth's total incoming solar energy (175,000 TW) and very low compared with our planet's total photosynthetic productivity (215 TW). The low redox disequilibrium value (=0.24 TW) is however likely to be broadly comparable with the energy associated with geochemical surface processes e.g. the dissolution of crust via precipitation on the modern Earth. Krissansen-Totton et al. (2016) compared atmospheric chemical redox disequilibria for Earth and other Solar System bodies and suggested that the large abundance of ($N_2$(g)-$O_2$(g)) in Earth's atmosphere together with the presence of global-scale oceans constitutes a significant source of thermodynamic disequilibrium. In general, there exists a wide range of abiotic processes known to generate redox disequilibrium - but their extent and magnitude on global scales compared with those due to life is not well-determined. Could, for example, the modern Earth's ($CH_4$(g)-$O_2$(g)) redox disequilibrium be produced abiotically? Today, up to 10% (equivalent to a few tens of Tg/yr) of total $CH_4$(g) emitted on Earth's surface is produced geothermally and the rest arises mainly via biology. On the early Earth, however, geothermal outgassing from $CH_4$(g) via mid-ocean ridges could have been significantly enhanced (Pavlov et al., 2000) compared with today. Also, regarding $O_2$(g), recent advances (see above) suggest that abiotic sources may have been strongly under-estimated depending on the planetary environment (instellation etc.). It seems therefore at least conceivable, that abiotic processes under certain circumstances could rival redox disequilibrium from biology.

## 2.1 $O_2$ combustion in Early Earth's atmosphere

Surface $O_2$ in early Earth's atmosphere reached a maximum abundance of ~0.3 bar during the Carboniferous period about (300-400) Myr ago likely via increased organic burial associated with widespread vascular land plant coverage (Dahl et al., 2010). Higher $O_2$ abundances were prevented however, likely due to $O_2$ combustion of organic carbon to form $CO_2$ as suggested by studies of fossilized-charcoal from paleofires initiated by lightning (Heath et al., 1999; Berner, 1999). Clearly there are differences between the case of early Earth (where large atmospheric oxygen abundances are driven mainly by life and where it is organic material which combusts) - compared with the proposed Super-Earth cases (where atmospheric oxygen is abiotically-formed and combustion takes place via the gas-phase oxidation of $H_2$(g)) . Nevertheless the case of early Earth is mentioned here as an example where $O_2$(g) is limited by combustion during a planet's evolution.

Although it may seem at first counter-intuitive, CE events early in a planet's history could in some ways favor the development of life e.g. firstly by helping to maintain oceans over longer timescales. Secondly, in ($O_2$-$N_2$) atmospheres combustion can form nitrogen oxides i.e. a form of fixed nitrogen useful for life. In ($O_2$-$CO_2$-$N_2$) atmospheres combustion can form hydrogen cyanide (HCN) (Giménez-López et al., 2010) which is a well-known precursor of amino acids (e.g. Yuasa et al., 1984).

### 3. Initiation of Combustion-Explosion in Planetary Atmospheres

Combustion and/or Explosion can be initiated when stable compounds such as molecular hydrogen are split by input of energy from lightning or/and cosmic rays to form reactive radicals. The resulting atoms can go on to initiate radical chain reactions which release energy faster than it can be removed by the surroundings.

### 3.1 Lightning

Similar to the way that an electric spark can be used in the laboratory to initiate combustion-explosion, lightning (and cosmic rays) represent natural phenomena which can also lead to these processes. An average single lightning flash delivers ~$6x10^9$ Joules on the modern Earth (Hill, 1991) and - analogous to an electrically-generated spark - leads to electrostatic discharge in the Earth's atmosphere.

Modern Earth features on average ~44 lightning flashes $s^{-1}$ (intra-cloud and cloud-to-ground combined) with generally more activity over land and in the tropics (Christian et al., 2003; Oliver, 2005). Earth's lightning activity breaks molecular nitrogen into atomic nitrogen – this reacts with oxygen compounds to likely produce (2-10)Tg (N)/year of nitrogen oxides (NOx) which can catalytically remove ozone in the stratosphere (e.g. Pickering et al., 1998).

On the early Earth, global lightning activity is not well constrained (Navarro-González et al., 1998; Navarro-González et al., 2001). Volcanic lightning was likely enhanced and thunderstorm lightning may also have been stronger due to possibly enhanced atmospheric dynamics due to a closer Moon and a faster rotating planet -  although more work to study such effects is required.

On Venus, lightning activity is estimated to be about 20% that of modern Earth (Russell et al., 2008) although optical evidence is still rather lacking (Cardesín-Moinelo et al., 2016; Yair, 2012). On Mars, electrical discharge is thought to occur frequently in dust devils and synoptic to global-scale dust storms (Yair, 2012 and references therein). On Jupiter and Saturn, integrated occurrence rates of lightning are estimated to be about one hundred times that of modern Earth and peak in the water cloud layers at 5 bar and 10 bar respectively (Yair, 2012 and references therein).

In summary, lightning is widespread in planetary atmospheres in the solar system. For SEs orbiting in the HZ of an M-dwarf star, General Circulation Model (GCM) studies (e.g. Joshi et al., 1997; Kite et al., 2011; Yang et al., 2013; Mills and Abbot, 2013) have suggested strong day-to-night circulation to maintain habitability which may provide wind velocities sufficient for charge separation hence favor the onset of lightning.

**3.2 Cosmic Rays**

Stellar (and Galactic) Cosmic Rays (CRs) **can** penetrate deeply into Earth's atmosphere (e.g. Veronnen et al., 2008). For SEs orbiting in the HZ of an active M-dwarf star, high inputs of Stellar and Galactic CRs could be present due to strong stellar activity, the close proximity to the star and the potentially weakened planetary magnetosphere associated with tidal-locking (e.g. Grieβmeier et al., 2005; Grenfell et al., 2007; Grenfell et al., 2012). Galactic Cosmic Rays with energies around the knee region of the energy spectrum and above i.e. $>10^{16}$ eV (about 1.6 mJ) could initiate (depending on atmospheric composition, T, p etc.) combustion. These have an occurrence rate near modern Earth's surface of several particles m$^{-2}$ year$^{-1}$. By comparison the "Minimum Ignition Energy" per spark to initiate combustion is ~0.03mJ (for $H_2$ in air), ~1mJ (for hydrocarbons in air) and ~1J for dust explosions in air (see also section 9) (Lackner, 2009).

**4. Combustive-Explosive Gas Mixtures of ($H_2$-$O_2$)**

Rapid release of energy can occur in gas mixtures when runaway chemical production (chain propagation) of free radicals occurs faster than the corresponding sink (termination) reactions which remove the free radicals. Depending on p, T there are in general two main mechanisms for energy release, namely via explosion (detonation) in which a pressure wave moves supersonically away from the ignition site and combustion (deflagration) in which a sub-sonic pressure wave together with electromagnetic radiation are generated. Combustion can occur either via energy input via sparks created when an applied electric field leads to dielectric breakdown of the gas molecules. Alternatively, combustion can occur via lightning or/and cosmic rays which can also lead to splitting or/and ionization of air molecules, or can be spontaneous, referred to as 'self-combustion'. The energy required to induce CE is termed the "minimum ignition energy" and is usually expressed in Joules.

Distinguishing between whether a given gas mixture explodes or combusts over a range of (p-T) is observationally challenging due to the power and complexity of the reaction mechanisms (see e.g. Sichel et al., 2002). Therefore in this work we use where possible the term "combustion-explosion (CE)" together. We now discuss CE for different gas-mixtures and place them in the context of exoplanetary atmospheres.

**4.1 Mechanism for [$H_2$-$O_2$] Combustion-Explosion**

Mixtures of $H_2$-$O_2$ gas, denoted as "oxyhydrogen", "electrolytic gas" or "detonating gas" are known to react either explosively or to combust, producing energy and the stable product water ($H_2O$). Assuming complete oxidation of $H_2$ by $O_2$, the overall (net) reaction is: $H_2 + \frac{1}{2}O_2 \rightarrow H_2O$. However, in practice the mechanism consists of intermediate steps in which other stable reaction products such as $H_2O_2$ can form. The key reaction steps of the $H_2$-$O_2$ CE mechanism (e.g. Cohen, 1992) are as follows:

$$\begin{align}
H_2 &\rightarrow H + H & (1) \\
H + O_2 &\rightarrow OH + O & (2) \\
O + H_2 &\rightarrow OH + H & (3) \\
OH + H_2 &\rightarrow H + H_2O & (4) \\
H + H + M &\rightarrow H_2 + M & (5) \\
O^{\#} + O + M &\rightarrow O_2 + M & (6) \\
O + H + M &\rightarrow OH + M & (7) \\
H + OH + M &\rightarrow H_2O + M & (8) \\
H + O_2 + M &\rightarrow HO_2 + M & (9) \\
HO_2 + H_2 &\rightarrow H_2O_2 + H & (10) \\
H/O/OH/HO_2 + surface^{\#\#} &\rightarrow products & (11)
\end{align}$$

[$^{\#}$O-atoms supplied into the system via e.g. $CO_2$ photolysis; $^{\#\#}$ removed from the system via gas-surface heterogeneous reactions occurring on the reaction chamber vessel or, in the case of a planetary atmosphere on the surface (if present) or on atmospheric aerosol].

Reaction (1) is the initiation step in which $H_2$ is dissociated in planetary atmospheres by lightning or by cosmic rays. $H_2$ has a minimum ignition energy in the range 0.02mJ/spark (US Dept. of Energy, Hydrogen Fact Sheet 1.008) to 0.03 mJ/spark (Lackner, 2009). This compares with a value of 0.29mJ for $CH_4$, with values generally >0.2mJ for higher hydrocarbons (Ono and Oda, 2008) and with values of typically ~1000mJ for dusts (many solids become very flammable when reduced to a fine powder in air). In general these energies depend on the gas composition, the total pressure and the spark duration (Maas and Warnatz, 1988; Ono and Oda, 2008). Note that the symbol 'M' in reactions (5) to (9) above refers to any third body present in the gas-phase required to remove excess vibrational energy of the reactants. Reactions (1-10) are commonly implemented in photochemical models in the literature. Required in order to simulate CE if it occurs are (i) the chemical heats of reaction which drive the rapid and runaway energy release, or/and (ii) treatment of cosmic rays which convert molecular into atomic hydrogen, or/and (iii) the energy budget via thermal diffusion, conduction etc. (see next section). Reactions (2)-(4) are the propagation steps.  Reactions (2) and (3) are called "chain branching" since they produce two reactive products namely (OH,O) and (OH,H) respectively from one radical reactant and can therefore lead to runaway propagation (production) of radicals. Reactions (5)-(9) represent the termination steps which overall remove chain carriers. Reaction (11) denotes sticky collisions of gas species with solid surfaces. Appendix one suggests that the importance of heterogeneous chemistry in removing reactive gas-phase species (reaction 11) is less for planetary atmospheres than for reaction vessels. This suggests that CE could be reached for a wider [p,T,composition] range in planetary atmospheres compared with reaction vessels.

Rapid release of energy occurs when the runaway propagation steps start to rapidly exceed the termination steps. Note that the mechanism produces $H_2O$ (reaction 8) and $H_2O_2$ (reaction 10) as stable products. With the addition of $N_2(g)$ i.e. on considering $(H_2-O_2-N_2)$ mixtures, there occur mixed nitrogen-oxygen reactions:

$$O + N_2 \rightarrow NO + N \qquad\qquad (12)$$
$$N + O_2 \rightarrow NO + O \qquad\qquad (13)$$
$$N + OH \rightarrow NO + H \qquad\qquad (14)$$

Reactions 12 and 13 are collectively referred to as the Zeldovich mechanism (Zeldovich, 1947).

## 4.2 Evolution of atmospheric [$H_2$:$O_2$] in Super-Earths

Figure 1 summarizes processes affecting $H_2(g)$ and $O_2(g)$ in SE atmospheres:

Figure 1: Processes affecting $H_2(g)$ and $O_2(g)$ in SE atmospheres.

In Figure 1, H-atoms from $H_2O$ photolysis can escape, especially during the active pre-main sequence phase of the star and can even drag off heavier O atoms during this stage. The remaining O-atoms (which can also be formed via $CO_2$ photolysis) can combine with themselves in a three-body gas-phase reaction to form $O_2(g)$ abiotically. $H_2(g)$ and $O_2(g)$ can either undergo CE if suitable conditions of [p,T, composition] are achieved to form water plus energy, or they can react via photochemistry. The amount of $H_2$ retained from the protoplanetrary disk therefore depends sensitively on the planet's mass, the size of the disk and the insolation from the star (Lammer et al., 2014; Luger et al., 2015) and can cover a wide range - from complete loss of $H_2$ up to about one percent $H_2$ of the total planetary mass (See also references in Table 1 below).

The amount of (abiotic) $O_2$ in the SE atmosphere is also predicted to cover a wide range depending on the UV from the central star and on model treatments of photolysis and atmospheric escape. Abiotic $O_2$ production proceeds e.g. via either carbon dioxide ($CO_2$) photolysis followed by recombination of oxygen (O) atoms with each other (e.g. Canuto et al., 1982) or, via water $H_2O$ photolysis followed by escape of atomic hydrogen (H) (Berkner and Marshall, 1964). The modern Earth features a column $O_2$ value of $4.5 \times 10^{24}$ molecules $cm^{-2}$ (Schneising et al., 2008). Segura et al. (2007) (their Table 2) however suggested abiotic $O_2$ amounts for $CO_2$-dominated atmospheres in the range $(2 \times 10^{18}$-$8 \times 10^{19})$ molecules $cm^{-2}$ (depending on the assumed outgassing rates, photochemical reaction rates, incoming UV etc.) i.e. up to about six orders of magnitude smaller than the modern Earth. Their study included rainout of oxidized species which led to a high abundance of reducing species (like $H_2$) hence their column $O_2$ values remained low. Model studies by Hu et al. (2012) (their Table 7) and Tian et al. (2014) (their Figure 3) - which included redox balance and thermal escape - suggested stronger abiotic $O_2$ amounts than the Segura study, i.e. about 100 times smaller than on modern Earth. Hu et al. (2012) calculated a mean mixing ratio over the atmospheric column of $1 \times 3 \times 10^{-3}$ $O_2$ for a terrestrial planet with a 90% $CO_2$ atmosphere orbiting a Sun-like star. Tian et al. (2014) suggested that the established OH-catalyzed cycles which drive the recombination of CO with O into $CO_2$ would be slow on SEs orbiting M-dwarf stars (i.e. favoring $O_2$ abiotic production up to 1000 times greater than for Sun-like stars) due to the weak Near-UV (NUV) output from the central star since NUV leads to release of atmospheric OH from its reservoirs (see also Harman et al., 2015). The model study by Domagal-Goldman et al. (2014) included redox balance of both the atmosphere and the ocean system and calculated modest abiotic $O_2$ columns of $(3 \times 10^{19}$-$2 \times 10^{21})$ molecules $cm^{-2}$ depending on stellar type, assuming atmospheres with 1bar surface pressure and $CO_2$ volume mixing ratios of 0.5. They suggested that model differences with the above-mentioned Hu and Tian studies could have arisen due to different treatments of CO removal from the atmosphere. Harman et al. (2015) noted the importance of redox balance in the atmosphere-ocean system and estimated an abiotic $O_2$ column of $9.3 \times 10^{22}$ molecules $cm^{-2}$ (i.e. ~2% of modern Earth) for an Earth-like planet with a 5%$CO_2$ atmosphere orbiting in the HZ of GJ876 (an M4V star). Wordsworth and Pierrehumbert (2014) suggested that planets with low abundances of non-condensing gases such as molecular nitrogen ($N_2$) would feature weak cold traps hence rapid $H_2O$ photolysis which could lead to efficient abiotic $O_2$ production. The presence of a large $H_2(g)$ atmosphere could therefore weaken or halt this mechanism although the location and magnitude of the cold trap in such atmospheres requires further investigation. Luger and Barnes (2015) modeled early stages of planets orbiting cooler stars (without large $H_2(g)$ envelopes) and suggested very large abiotic $O_2$ with up to two thousand times the mass of Earth's atmospheric $O_2$. In their study abiotic production is favored by strong incoming X-ray Ultra Violet (XUV) radiation from young (up to 1Gyr) pre-main sequence M-dwarf stars which drives fast photolysis of $H_2O$ and escape of the resulting H in the planetary atmosphere. Regarding spectral features, Schwieterman et al. (2016)[a,b] discuss possible means of identifying abiotic $O_2$ spectral signals. García Muñoz et al. (2009) investigated spectroscopic features of the $O_2$ dimer nightglow. More work is required to constrain better the range of possible $CO_2$ and $H_2O$ amounts from outgassing (e.g. Lammer et al., 2013) available to form $O_2$ abiotically. Table 1 summarizes the total mass range of $H_2(g)$ and $O_2(g)$ estimated from the literature to occur in SE atmospheres (note also the caveats discussed below):

| Quantity | Value | Reference |
|---|---|---|
| Mass Earth ($M_e$) (g) | $5.97 \times 10^{27}$ | NASA Earth fact sheet 2017 |
| Mass Earth Atmosphere (g) | $5.10 \times 10^{21}$ | as above |
| Mass SE with $2r_e$[§] | $4.78 \times 10^{28}$ | |
| $H_2(g)$ in SE atmosphere | $(2.5 \times 10^{19} - 1.5 \times 10^{26})$<br>$4.78 \times 10^{26}$<br>$3.57 \times 10^{23}$ | Lammer et al. (2014)[*]<br>Chiang and Laughlin (2013)[#]<br>Miller-Ricci and Fortney (2010)[§§] |
| $O_2$ (g) in modern Earth atmosphere | $1.07 \times 10^{21}$ | NASA Earth fact sheet 2017 |
| Abiotic $O_2$ (g) in SE atmosphere[#] | $2.38 \times 10^{24}$<br>$(2 \times 10^{18} - 8 \times 10^{19})$ | Luger and Barnes (2015)[**]<br>Segura et al. (2007)[+] |
| $O_2$ (g) in Archaean –type atmosphere | $5.67 \times 10^{16}$ | Gebauer et al. (2017)[##] |

Table 1: Range of total atmospheric masses of $H_2(g)$ and $O_2(g)$ for Super-Earths calculated from the literature. [§]Assuming SE with Earth's density. [*]Model of nebula gas accretion - range shown depends on assumed dust loadings of nebula, stellar luminosity etc. [#]Assuming $H_2(g)$ constitutes 1% of total SE mass. This work suggested that the assumed planetary scenario would accrete dry which implies a limited potential for abiotic $O_2(g)$ formation (see also text to Table 1). [§§]For GJ1214b assuming $H_2(g)$ constitutes 0.05% of the planet's mass, or ~70 times the mass of Earth's atmospheric oxygen. [**]Due to strong water photolysis and subsequent escape of H during the pre-main sequence phase. This work quotes up to 2000 times the mass of Earth's atmospheric oxygen. [+]Due to $CO_2$-photolysis. [##]Assuming $10^{-5}$ times the modern atmospheric level of $O_2(g)$.

In Table 1, the $H_2(g)$ mass range varies from ~($10^{19}$-$10^{26}$)g whereas $O_2(g)$ varies from ~($10^{16}$-$10^{24}$g). Estimating however which combinations of $H_2(g)$ and $O_2(g)$ could actually co-exist in SEs in nature, requires further studies with coupled chemistry-climate models which calculate consistent, gas-phase chemical evolution. Note that the extent to which SEs in the HZ of low mass stars accrete water during their formation (hence their ability to form $O_2(g)$ abiotically via water photolysis followed by H-escape) is a subject of discussion (see e.g. Lissauer, 2007; Hansen, 2015). Recent model studies addressing abiotic $O_2(g)$ formation (e.g. Hu et al., 2012; Tian et al., 2014; Harman et al., 2015) all assume low reducing conditions with relatively small $H_2(g)$ abundances.  It is challenging for current coupled climate-photochemistry models to operate over such large p, T and composition ranges including all relevant processes (climate, chemistry, escape etc.)  Notable model studies in this direction, however are e.g. Miller-Ricci and Fortney (2010); Miguel and Kaltenegger (2013); Seager et al., (2013) and Hu and Seager (2014). Similarly it is challenging to simulate thick steam atmospheres associated with the strongest abiotic $O_2(g)$ production scenarios. The potentially large range of abiotic sources of $O_2(g)$ shown in Table 1 are often-quoted in the exoplanetary community often without reference to photochemical sinks of $O_2(g)$ due to large $H_2(g)$ envelopes which could be present in SEs.

Figures 2a-c show schematically examples of the evolution of key chemical species in SE atmospheres for fast (molecular) hydrogen loss (Figure 2a), medium hydrogen loss (Figure 2b) and slow hydrogen loss (Figure 2c):

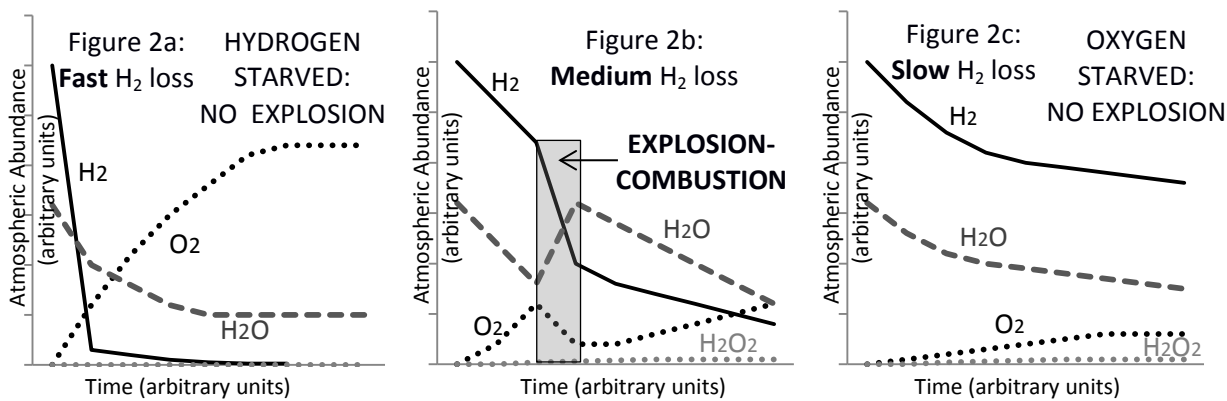

Figure 2: Schematic evolutionary pathways of key chemical species in SE atmospheres for three hypothetical cases, namely fast $H_2$ loss (Figure 2a, left panel), medium $H_2$ loss (Figure 2b, middle panel) and slow $H_2$ loss (Figure 2c, right panel). Hydrogen loss can be driven by either strong incoming EUV or a large amount of hydrogen initially accreted. The shaded rectangular region in Figure 2b shows species' response to possible combustion-explosion. Species shown are: $H_2$ (solid black line), $H_2O$ (dashed black line), $O_2$ (dotted black line) and $H_2O_2$ (dotted grey line).

The three panels in Figure 2 describe hydrogen loss as well as oxygen and water formation under different EUV conditions. Hydrogen loss depends on e.g. the initial amount of $H_2$ accreted from the protoplanetary disc and upon the escape rate of hydrogen atoms. This escape is essentially driven by EUV insolation and can proceed in two steps – first via photochemical release of hydrogen atoms from hydrogen-containing molecules (e.g. $H_2O$, $CH_4$, $H_2$ etc.) (which takes place in the middle atmosphere and above) and second via diffusion- or/and energy-limited escape. Oxygen (abiotic) formation proceeds via gas-phase combination of O-atoms which originate from water photolysis or from photolysis of other O-containing species such as $CO_2$.

Figure 2a shows the "$H_2$-starved case" i.e. where the CE limit is not reached because hydrogen loss is rapid (associated with strong incoming EUV) so that its abundance drops below the CE limit before $O_2$ can sufficiently build-up. Figure 2b shows the case where hydrogen is lost more slowly than in Figure 2a hence the system remains above the CE $H_2$-lower limit for longer which means more time for build-up of abiotic $O_2$. Then, the CE limit can be reached as denoted by the grey-shaded rectangle. This results in rapid ocean formation after a phase of ocean loss via evaporation and photolytic dissociation of water. In Figure 2b assumes that the water reservoir is never larger than the initial inventory. In addition to forming $H_2O$, CE in Figure 2b leads to formation of some $H_2O_2$ which is gradually lowered e.g. associated with the photolytic loss of $H_2O$. Figure 2c shows the "$O_2$-starved case" i.e. where the CE limit is now not reached because weak incoming EUV leads to insufficient build-up of $O_2$.

The effects of (classical, gas-phase) photochemistry should also be considered in Figure 2 in addition to CE. For example, high $H_2$ and high EUV could lead to faster photochemical removal of $O_2$ by $H_2$ which could prevent $O_2$ from reaching the CE limit. These effects should be the focus of future work which requires a wider parameter range of study than possible in the present work.

## 4.3 Temperature-pressure dependence of [$H_2$-$O_2$] explosion

Figure 3 shows the characteristic S-curve for the [T-p] dependence of [$H_2$-$O_2$] explosion:

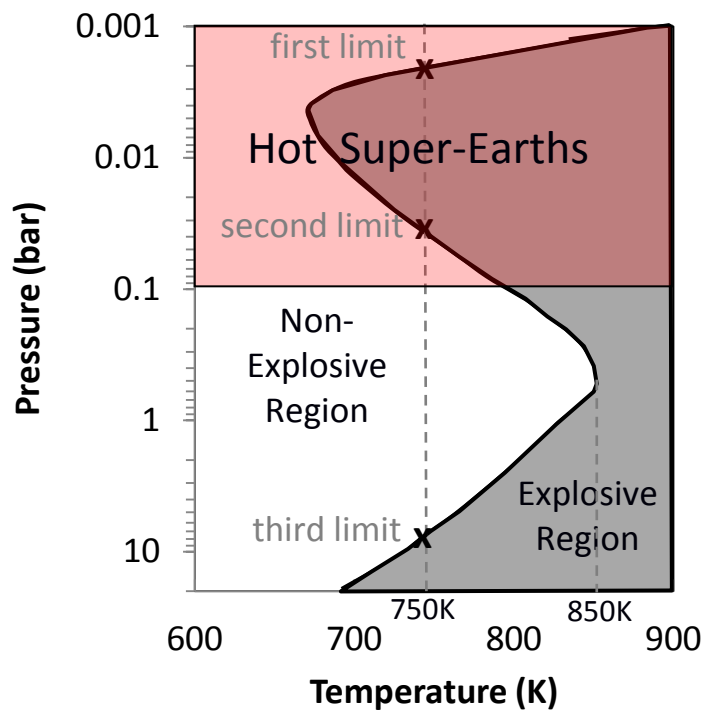

Figure 3: Temperature-pressure dependence of ($H_2$-$O_2$) explosion. Data source adapted from Lewis and von Elbe (1987) for a two-to-one hydrogen-to-oxygen stoichiometric mixture using a spherical vessel 7.4cm in diameter with a potassium chloride coating. The explosive region is shaded in grey, the non-explosive region is non-shaded. As an example at T=750K, the three points marked as "X" along the grey dashed line denote the first, second and third explosive limits i.e. where the grey-shaded and non-shaded regions cross. The red-pink shaded rectangle in the upper part of the Figure shows the relevant range (0.1-0.001bar) sampled via transit transmission spectroscopy (see e.g. Hu and Seager, 2014).

Figure 3 shows the case for a 2:1 [$H_2$:$O_2$] stoichiometric mixture (although the CE limit can be attained over a range of [$H_2$:$O_2$] mixtures depending on p, T, see Figure 4 below). Note too that for the particular stoichiometric composition (H:O=2:1) assumed in Figure 3, CE proceeds for only ~T>660K. Note that achieving the same stoichiometry as assumed in Figure 3 combined with these relatively high middle atmosphere temperatures might therefore be limited to the early stages of a planet's evolution if the thick $H_2$-dominated atmosphere is permanently lost thereafter. SEs however likely cover a wide [p,T,composition] range as discussed below e.g. from potentially habitable conditions such as recently suggested for Kepler 452b (Jenkins et al., 2015) to the hot, thin atmospheres of SEs such as CoRoT 7b (Hatzes et al.,2011) where surface T at the sub-stellar point likely exceeds 2000K. The white (non-explosive) and grey (explosive) regions in Figure 3 can be interpreted as follows:

At pressures lower than the first limit in Figure 3, the mixture is non-explosive (corresponding to the unshaded region at the top of Figure 3) due to efficient diffusion favoring wall-reactions on the reaction vessel (for an estimation of this effect for atmospheres, see **appendix 1**) which remove reactive radicals. On increasing pressure, diffusion slows and the mixture becomes explosive i.e. the rate of the propagation reactions exceed that of the termination reactions - at the 'first (explosive) limit'. On increasing pressure further, the mixture becomes once more non-explosive at the 'second limit' because the pressure-dependent reaction 9 (whose rate varies approximately with $p^2$) is now important in removing H atoms; Lee and Hochgreb (1998) discuss effects affecting the second explosion limit and present possible chemical pathways for $H_2$ oxidation. At higher pressures still, reaction 10 can become important in producing H and the mixture becomes once more explosive at the 'third limit'. Schroeder and Holtappels (2005) present the lower and upper explosion limits shown in the number of moles of $H_2$ present as a % of the total moles (mol% $H_2$) as a function of pressure. The lower limits vary from 4.3% mol% $H_2$ (1 bar) up to 5.6% mol% $H_2$ (150bar); the upper limits vary from 76.5% mol% $H_2$ (1 bar) (which corresponds to a lower limit of 23.5% $O_2$) down to 72.9% mol% $H_2$ (150bar). Zheng et al. (2010) suggested that experimental design (chamber size, shape, wall-coating etc.) leads to an error in the derived (p-T) of both the lower and upper explosion limits by about 4%.

At temperatures above about 850K (see Figure 3) the system is explosive for all pressures. This is because propagation reactions have generally moderately positive temperature-dependencies whereas termination reactions have either only weakly positive or weakly negative temperature dependencies. At intermediate temperatures (700-850K) the system can be explosive or not depending on the pressure. Maas and Warnatz (1998) provide more details on the T-dependence of propagation and termination reactions. Schroeder and Holtappels (2005) present the lower and upper explosion limits (in mol% $H_2$) as a function of temperature. The lower limits vary from 3.9% (293K) down to 1.5% (673K); the upper limits vary from 75.2% (293K) up to 87.6% (673K).

**4.4 Composition dependence of [$H_2$-$O_2$] combustion**

Figure 4 shows the combustion (flammability) regions for ($H_2$-$O_2$-$N_2$) and for ($H_2$-$O_2$-$CO_2$) gas mixtures at T=298K and p=1 bar:

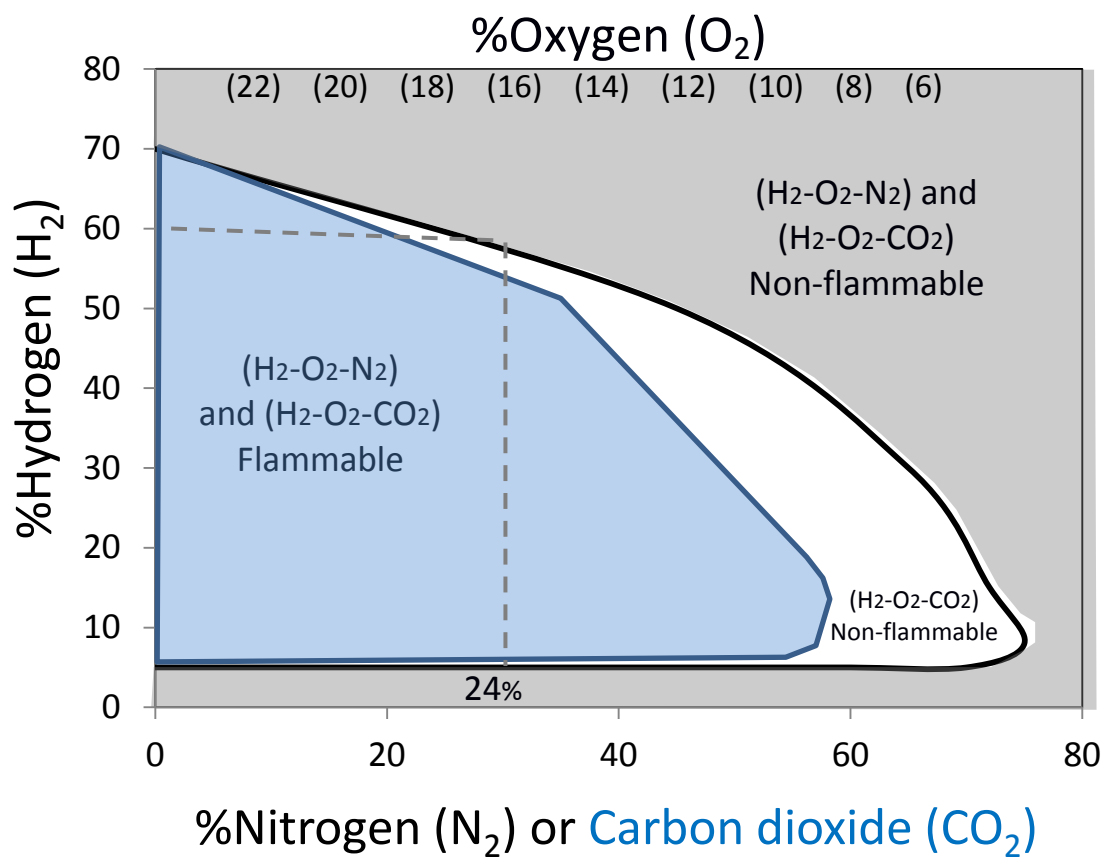

Figure 4: Compositional dependence of the (H$_2$-O$_2$-N$_2$) and (H$_2$-O$_2$-CO$_2$) systems (shown in molar concentration by percent) upon combustion for gas mixtures at T=298K, p=1bar. The Figure shows %H$_2$ concentration (y-axis) and %N$_2$ (or %CO$_2$) concentration (x-axis) with the remaining ("leftover gas") being O$_2$. In the shaded blue region both (H$_2$-O$_2$-N$_2$) and (H$_2$-O$_2$-CO$_2$) mixtures are flammable. In the shaded grey region both mixtures are non-flammable. In the central white region (H$_2$-O$_2$-CO$_2$) mixtures are non-flammable whereas (H$_2$-O$_2$-N$_2$) mixtures are flammable. The dashed grey line shows as an example the %molar gas composition of [60:16:24] for [H$_2$:O$_2$:N$_2$]. Data shown is adopted from the same source as for Figure 3.

Figure 4 suggests that (H$_2$-O$_2$-N$_2$) mixtures at T=298K and p=1bar are combustive (flammable) for H$_2$ concentrations of about (5-70%). Other studies (Schroeder and Holtappels, 2005) reported similar limits at these (p,T) i.e. suggesting a lower limit of (3.6-4.2%)H$_2$ and an upper limit of (75.1-77.0%). Cohen (1992) on the other hand, (their Table 1 and references therein) suggested e.g. for O$_2$/N$_2$ of (21:79) an %H$_2$ lower limit from (4.2-9.4)% and upper limit from (64.8-74.7)%. In short, above about 70% N$_2$, the mixture in Figure 4 is non-combustive, whereas for high H$_2$ concentrations this value decreases to ~0% N$_2$. On increasing the temperature for (H$_2$-O$_2$-N$_2$) mixtures (e.g. to 300$^{\circ}$C, not shown) the % lower (upper) combustive limit for H$_2$ concentration in Figure 4 is lowered (raised) by a few percent.

What is the effect of changing the background gas from molecular nitrogen to other inert gases? For ($H_2$-$O_2$-$CO_2$) mixtures, Figure 4 suggests that the "nose feature" at ~10% $H_2$ is shifted to the left compared to the ($H_2$-$O_2$-$N_2$) case with the crossover between flammability and non-flammability for ($H_2$-$O_2$-$CO_2$) occurring at ~60% $CO_2$. Cohen (1992) (again their Table 1 and references therein) suggested for $O_2/CO_2$ (21:79) an %$H_2$ lower limit from (5.3-13.1)% and upper limit from (68.2-69.8)%. Changing the background gas from $N_2$ to helium (He) (Cohen, 1992) leads to STP flammability limits of 7.8% $H_2$ (lower limit) and 75.7% $H_2$ (upper limit) for a molar ratio of ($O_2$/He) similar to air but where helium replaces nitrogen.

Particularly relevant is to consider the effect upon combustion-explosion of a background steam atmosphere. This is because CE relies on rapid build-up of abiotic $O_2$. This, in turn requires the presence of sufficient steam in the early stages after planet formation to drive water photolysis followed by hydrogen escape. Changing the background gas from $N_2$ to steam leads to an increase in OH, a reduction in NO and possibly an increase in flame temperature according to the study by Park et al. (2004). Processes which favor the conversion of OH into $HO_2$ would likely favor $H_2O_2$ formation since $HO_2$ is a major in-situ source of hydrogen peroxide. Singh et al. (2012) performed a modeling study of syngas combustion in air which suggested increased abundances of H, OH and $HO_2$ on increasing the steam content.

## 5. Model Studies of [$H_2$-$O_2$] Atmospheres

### 5.1 Motivation and Aim

It is challenging for the current generation of 1D coupled convective-climate-chemistry models to cover the potentially large pressure, temperature, and atmospheric composition e.g. ($H_2$-$O_2$) or **mass** (up to hundreds of bar) range predicted for some SEs and studies thereof are rather lacking in the literature. How these species interact and evolve can affect habitability in different ways. For thick, $H_2$-dominated atmospheres Rayleigh scattering hence surface cooling can become important. Also, a decreased molecular weight leads to an increased atmospheric scale height. The presence of $H_2$(g) enhances pressure-broadening which increases greenhouse gas efficiencies. In addition to such effects for $H_2$(g), the amount of (abiotic) $O_2$(g) in SE atmospheres is clearly also relevant for interpreting false positives in biosignature science.

The aim of our model study here is to investigate the mechanism and timescales of [$H_2$- $O_2$] oxidation in SE atmospheres. We estimate thereby the chemical pathways, the timescales over which the standard gas-phase chemistry affects $H_2$(g) and $O_2$(g) and the consequences for CE. We investigate here only a modest part of the expected parameter range in order to remain in the region where our 1D climate-chemistry model (see description below) is valid. Future work plans to extend the model chemical network to simulate thick, primary and steam atmospheres.

## 5.2 Model Descriptions

### 5.2.1 Atmospheric Column Model

The cloud-free 1D stationary model applied here consists of an atmospheric (convective-climate-photochemical) module and a biogeochemical module as described in Gebauer et al. (2017). The climate scheme uses updated longwave and shortwave modules described in von Paris et al. (2015). The incoming shortwave (0.2-4.5 microns) scheme employs 38 bands with a two-stream approach from Toon et al. (1989) and with Rayleigh scattering parameterizations included for $N_2$, $H_2$, $H_2O$, He, CO, $CO_2$ and $CH_4$ (Shardanand and Rao, 1977; von Paris et al., 2015). The longwave scheme (1-500 microns) employs 25 bands for molecular absorption by $H_2O$, $CO_2$, $O_3$ and $CH_4$. The atmospheric module extends from the ground up to the mid-mesosphere for the modern Earth, assumes the Earth's biomass and development, and consists of two main components: firstly, a photochemical module and secondly, a convective-climate module. The original chemical module was described in Kasting et al. (1984) with updates and validations as described in Gebauer et al. (2017). The climate module assumes convective adjustment in the lower atmosphere. In the middle atmosphere and above the scheme solves the radiative transfer equation including a parameterization for incoming shortwave radiation, outgoing longwave radiation for the major absorbers as described in von Paris et al. (2015).

### 5.2.2 Pathway Analysis Program

The Pathway Analysis Program (PAP) (Lehmann, 2004) was applied to reaction rates and concentrations output over consecutive timesteps output from the 1D atmospheric model. PAP is a useful diagnostic tool for identifying and quantifying potentially complex chemical pathways in planetary atmospheres – in this case the pathways relevant for oxidation by $O_2(g)$ of $H_2(g)$. When building the pathways step-by-step, the PAP algorithm discarded pathways below the user-set minimum flux ($f_{min}$) of $10^{-11}$ vmr/s $O_2(g)$ in order to avoid combinatorial explosion (see Lehmann , 2004).

**5.3 Scenarios**

RUN 1 is the modern Earth control. Surface fluxes of key source gases and biomass emissions were adjusted to reproduce modern Earth's global mean surface atmospheric volume mixing ratios (vmr) as described in Gebauer et al. (2017) with values $O_2$=0.21, Ar=0.01, $CO_2$=3.55x10$^{-4}$, $CH_4$=1.6x10$^{-6}$, $N_2O$=3.0x10$^{-7}$, $CH_3Cl$=5x10$^{-10}$ vmr. $H_2$ at the surface was set to a constant value of 5.5x10$^{-10}$. $N_2$ (~0.78 vmr) was a fill gas such that the total surface pressure reached one atmosphere. Surface albedo was fixed to a value of 0.212 in order to reproduce Earth's global mean surface temperature of 288K.

RUN 2 is as for run one but for a Super-Earth with x3 Earth's gravity and x1000 surface $H_2$ (=5.5x10$^{-4}$) ("SE x1000$H_2$ run") assumed to have one third of Earth's atmospheric mass (so that $P_{surf}$=~1bar). This mass of $H_2$ corresponds to about four tenths of a percent of the total mass of the SE atmosphere. Run two therefore simulates an SE with a small-to-modest amount of $H_2$(g) left over from accretion, or a planet with strong $H_2$(g) geological sources or/and atmospheric in-situ sources. All other planetary and stellar input parameters are set to modern Earth values as described in Gebauer et al. (2017). Note that future model development is required to simulate higher $H_2$ abundances.

**6. Model Results**

**6.1 Temperature**

Figure 5 compares the temperature (K) profiles for run 1 (modern Earth, solid line) and run 2 (3g SE with x1000 increased $H_2$, dotted line). In the mid-stratosphere and above strong cooling of up to ~40K occurs in run 2. This is related firstly, to increased $CH_4$(g) absorption (see Figure 6) since the enhanced $H_2$ led to a decrease in OH (a strong methane sink) via the reaction between ($H_2$+OH) and secondly, due to increased Rayleigh scattering in the enhanced $H_2$ atmosphere. In the troposphere there occurred modest overall cooling in run 2 (despite enhanced $CH_4$(g)) by up to a few degrees. This arose because firstly, run 2 (with 3g, $P_o$=1bar x1000$H_2$) has ~one third of Earth's atmospheric mass of run 1 hence a weaker overall greenhouse effect and secondly, due to enhanced Rayleigh scattering from increased $H_2$(g).

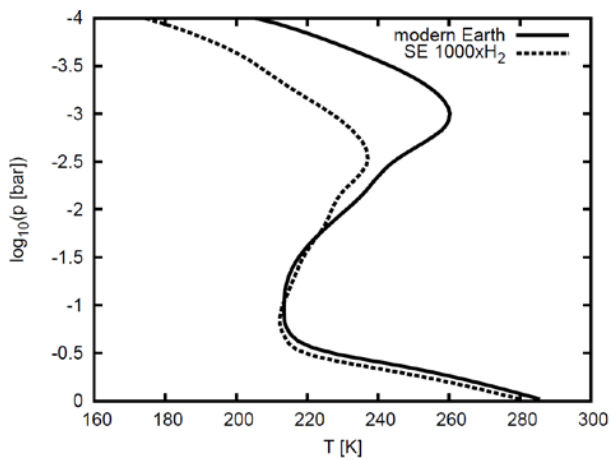

Figure 5: Modelled temperature (K) profile for run 1 (modern Earth, solid line) and run 2 (3g SE with x1000 increased H$_2$, dotted line).

## 6.2 Chemical Abundances

Figure 6 compares key chemical abundance profiles for run 1 (modern Earth) and run 2 (3g SE with x1000 increased H$_2$(g)).

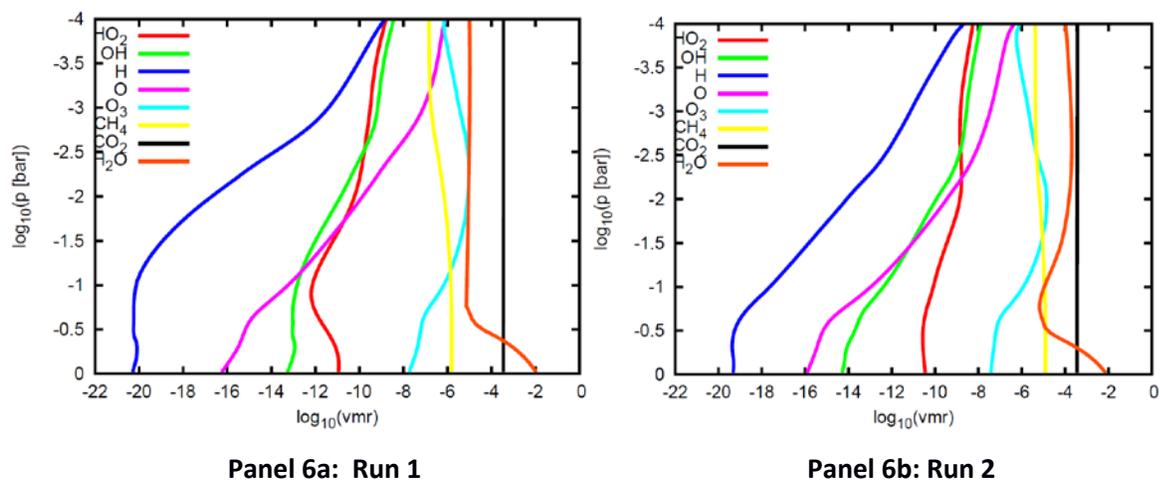

**Panel 6a:  Run 1**                    **Panel 6b: Run 2**

Figure 6: Modelled chemical abundance (vmr) profiles for run 1 (modern Earth) (Figure 6a, left panel) and run 2 (3g SE with x1000 H$_2$) (Figure 6b, right panel).

Figure 6 suggests an OH reduction in run 2 (Figure 6b) (hence enhanced CH$_4$(g)) as already discussed. Ozone and atomic oxygen profiles are broadly similar in both runs. Responses in tropospheric water (mainly driven by changes in temperature which drive evaporation and condensation in this region) are also rather small due to rather weak tropospheric temperature changes (see Figure 5 and discussion above). Table 2 shows atmospheric column values in Dobson Units for key chemical species.

| Species | Column (DU) Modern Earth (run 1) | Column (DU) Super Earth x1000 $H_2$ (run 2) |
|---|---|---|
| Ozone ($O_3$) | 305 | 167 (102) |
| Methane ($CH_4$) | 1231 | 3082 (410) |
| Water ($H_2O$) | $2.4 \times 10^6$ | $5.0 \times 10^5$ ($8.0 \times 10^5$) |
| Chloromethane ($CH_3Cl$) | 0.36 | 1.8 (0.12) |

Table 2: Column values (Dobson Units [DU]), 1DU=$2.69 \times 10^{16}$ molecules cm$^{-2}$) for key chemical species. Grey values show response of chemically insert species due to reducing the atmospheric column by x3 for a 3g SE (see main text).

Grey bracketed values in the far right-hand side column of Table 2 show one third of the modern Earth (run 1, middle column) value i.e. the value which a chemically-inert species would have if the total atmospheric mass in run 1 is reduced by a factor of three. Differences between the black and grey values in the right-hand side column therefore arise due to e.g. photochemistry and temperature responses.

In the far right column of Table 2, ozone is increased for the black value (with chemistry) compared to the grey value (without chemistry). This arose firstly due to the reduction in OH in the middle atmosphere (as discussed) (favouring more ozone) and secondly due to mid stratosphere cooling (see Figure 5) which slowed the Chapman sink reaction between ($O_3$+O) hence led to ozone production. Methane and chloromethane in the far right column of Table 2 have enhanced values for the values written in black (with chemistry) compared with the values written in grey (without chemistry). This strong effect is related to the decreased OH in run 2 as discussed above. OH is an important sink for these species especially in the troposphere where most of the column resides and changes in OH can lead to non-linear responses in these species' concentrations. The water value shown in black (no chemistry) is somewhat lower than the grey value due to tropospheric cooling (hence enhanced condensation) in run 2.

Table 3 shows material fluxes of hydrogen and oxygen atoms (teragrammes/yr) at the uppermost model boundary.

| Species | This Study (Tg/yr) [Bar/Myr] | | Luger and Barnes (2015)* [Bar/Myr] |
|---------|--------------|--------------|-----------------------|
| H | Run 1 = (-0.2) | | - |
|   | Run 2 = (-4.6) | | - |
| O | Run 1 = (+2.2) | [+2.1x10$^{-4}$] | [+25]* |
|   | Run 2 = (+1.3) | [+1.3x10$^{-4}$] | |

Table 3: Material fluxes (Tg/yr) across the model upper boundary. Positive values indicate removal (upwards escape) whereas negative values indicate input (downwards effusion) into the model domain. *Model study calculating accumulated abiotic atomic oxygen due to water photolysis followed by hydrogen escape for a SE orbiting in the inner HZ of an M-dwarf star during the pre-main sequence.

In Table 3 (far right column) the bar unit refers to an atmospheric column with modern Earth's (1g) mass and composition which corresponds to 0.21 bar of diatomic oxygen. H-escape fluxes ($\phi_H$) in Table 3 are calculated from the diffusion-limited formula based on Walker (1977):

$$\phi_H = 2.5x10^{13} [f_{total}] \; H \qquad\qquad (15)$$

where $f_{total}$ denotes sum of hydrogen-containing species abundances in the uppermost model layer, H=atmospheric scale height. O-fluxes ($\phi_O$) in Table 3 represent the downward flux which arises at the model lid due to photolysis of $CO_2$ (Segura et al., 2003) calculated via:

$$\phi_O = jCO_2 \; [CO_2] \; H \qquad\qquad (16)$$

where $jCO_2$ is the photolyis coefficient of $CO_2$ and $[CO_2]$ denotes the $CO_2$ abundance in the uppermost model layer. Material fluxes shown in Table 3 for this work are quite modest - as one would expect for conditions which do not vary greatly from modern Earth where our model is valid. By comparison the fluxes for the extreme conditions calculated by Luger and Barnes (2015) (grey values Table 3) are much stronger. Future work (see also discussion) will apply a new model version currently under development for $H_2$-dominated atmospheres with stronger hydrogen and oxygen material fluxes.

**6.3 Chemical Production and Loss**

Figures 7a and 7b shows the difference (production – destruction) in the net gas-phase reaction rates of $O_2$ (g) (Figure 7a) and $O_3$(g) (Figure 7b). For the Earth control (run 1, red line), Figure 7a suggests modest $O_2$(g) chemical loss peaking in the mid-stratosphere at ~40km and modest production peaking in the upper stratosphere at ~50km. For the SE (run 2, blue dashed line), Figure 7a suggests a stronger response with $O_2$(g) loss at (15-18km) and $O_2$(g)production above 18km. How does one interpret the two regions of net chemical production and loss? In our column model, chemical concentrations converge to steady-state. In other words the net result of gas-phase chemistry, transport, emission, deposition etc. is zero. In Figure 7a, the mid-stratosphere region with net chemical loss is balanced by transport of $O_2$(g) via Eddy diffusion into that region - and vice-versa for the upper-stratosphere region with net chemical production.

Figure 7b is as for Figure 7a but for $O_3$(g). One sees that Figures 7b and 7a are approximately mirror-images of each other. This suggests that in-situ photochemistry leads to the interconversion of $O_2$(g) and $O_3$(g) over altitude. For the modern Earth (run 1, red line) for example, there is net chemical loss of $O_2$(g) into $O_3$(g) in the mid-stratosphere where the ozone layer peaks – and vice-versa in the upper stratosphere.

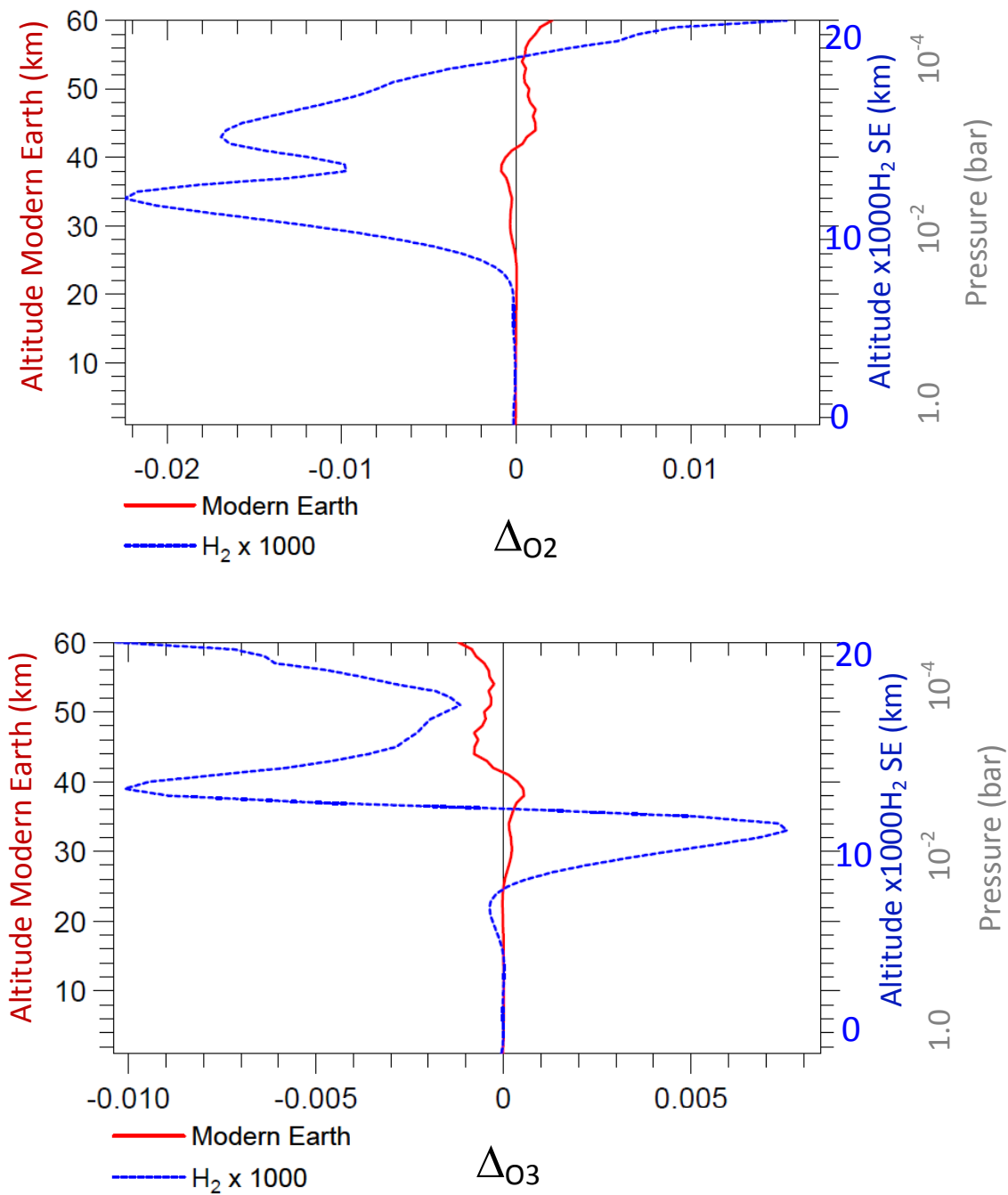

Figure 7: Difference (production minus loss) for atmospheric in-situ gas-phase rates in (ppbv/s) for oxygen ($\Delta O_2$) (Figure 7a) (upper panel) and for ozone ($\Delta O_3$) (Figure 7b) (lower panel) for the modern Earth (run 1, red continuous line) and the SE x1000H$_2$ (run 2, blue dashed line). Note that the top four model layers (corresponding to 61-64km for the Earth control, run 1) are omitted due to model boundary effects in the upper lid.

Performing a pathways analysis of O$_2$(g) in runs one and two therefore leads to the construction of pathways converting O$_2$(g) into O$_3$(g) and vice-versa. These pathways are rather complex and are mostly HOx-catalysed. In the framework of this paper however, the focus is not upon interconversion pathways of $O_2(g)$ and $O_3(g)$ (not shown), but instead on the reduction of $O_2(g)$ by $H_2(g)$ to form $H_2O$ or/and $H_2O_2(g)$. In order to analyse this latter process, one can define the "Oy" family, where: Oy=$[2O_2+3O_3+O(^3P)+O(^1D)+OH+2HO_2+2H_2O_2+2ClO_2+ClO+NO+2NO_2]$. Performing a pathway analysis for Oy will therefore not consider conversions between shorter-lived members of the oxygen family. It shows instead chemical pathways e.g. for the net reaction: $O_2+2H_2 \rightarrow 2H_2O$. Figure 8 is as for Figure 7 but for the Oy family:

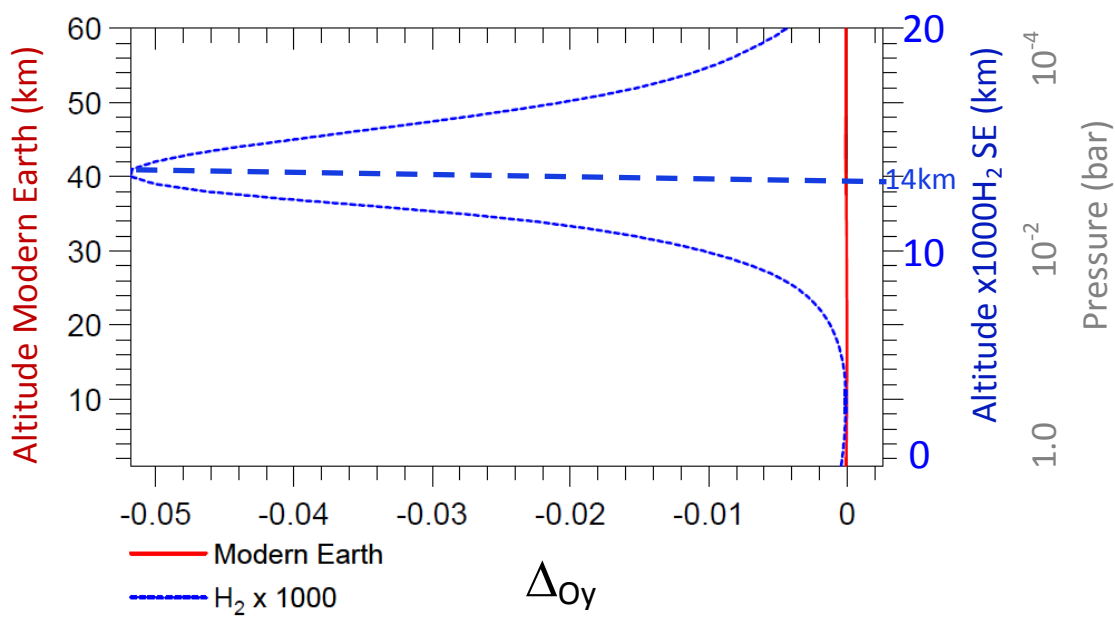

Figure 8: Difference (production minus loss) for atmospheric in-situ gas-phase rates of change in (ppbv/s) for the "Oy" family ($\Delta$Oy) where Oy=$[2O_2+3O_3+O(^3P)+O(^1D)+OH+2HO_2+2H_2O_2+2ClO_2+ClO+NO+2NO_2]$. Results are shown for the modern Earth (run 1, red continuous line) and the SE x1000$H_2$ (run 2, blue dashed line).

Figure 8 suggests that for the modern Earth (run 1, red line), in-situ gas-phase changes in Oy are close to zero over the altitude range considered. This means that although the concentrations of Oy family members can interchange over altitude (e.g. some $O_2$ is converted into $O_3$ in the stratosphere), the overall concentration of Oy is conserved. For the SE x1000$H_2$ run (run 2, blue dashed line) there is a distinct peak in the most negative values of $\Delta$Oy at ~14km. Why? This occurs mainly due to removal via the net oxidation reaction: $2H_2 + O_2 \rightarrow 2H_2O$ (see PAP analysis below) which proceeds via HOx catalysed pathways. Below 14km HOx concentrations are low and the rate of the oxidation reaction (hence the deviation of $\Delta$Oy away from zero) is negligible. At higher altitudes >~20km, although $H_2O$ is formed via the oxidation reaction, it is then photolysed rapidly into HOx which means no overall effect upon Oy (see Oy definition above). In summary, gas-phase reactions which are responsible for ($H_2$-$O_2$) oxidation operate mainly within a narrow band in the middle atmosphere.

## 6.4 Pathway Analysis

A pathway analysis was performed for Oy in run 2 (x1000$H_2$ SE) in order to determine the main pathways for ($H_2$-$O_2$) removal and to estimate removal timescales based on gas-phase mass fluxes through the pathways found. The analysis was performed in the region where ($H_2$-$O_2$) oxidation is most effective i.e. at ~14km where $\Delta$Oy reaches its most negative value in Figure 8. Calculations were based on two consecutive timesteps of converged atmospheric column model output over which $\Delta$Oy=-21.22 ppt/s. Results are shown in Table 4 for all pathways which individually contribute >1% to the overall flux ($\Delta$Oy) over the interval analysed. These pathways collectively account for 86.3% of the total removal rate of Oy in the model in this layer. The remaining 13.7% is attributable to minor pathways which individually contribute less than 1% (not shown).

| Pathway | %Loss* | Comments |
|---------|--------|----------|
| Pathway 1<br>$O_2 + h\nu \rightarrow O(^3P) + O(^3P)$<br>$2[O(^3P) + HO_2 \rightarrow OH + O_2]$<br>$2[OH + H_2 \rightarrow H_2O + H]^{\#}$<br>$\underline{2[H + O_2 + M \rightarrow HO_2 + M]}$<br>net: $O_2 + 2H_2 \rightarrow 2H_2O$ | 47.8% | Oxidation of $H_2$ into $H_2O$ catalysed by HOx |
| Pathway 2<br>$O_2 + h\nu \rightarrow O(^3P) + O(^3P)$<br>$2[O(^3P) + O_2 + M \rightarrow O_3 + M]$<br>$2[O_3 + h\nu \rightarrow O_2 + O(^1D)]$<br>$2[H_2 + O(^1D) \rightarrow OH + H]^{\#}$<br>$2[H + O_2 + M \rightarrow HO_2 + M]$<br>$\underline{2[OH + HO_2 \rightarrow H_2O + O_2]}$<br>net: $O_2 + 2H_2 \rightarrow 2H_2O$ | 17.1% | Oxidation of $H_2$ into $H_2O$ catalysed by HOx and involving $O_3$ |
| Pathway 3<br>$O_2 + h\nu \rightarrow O(^3P) + O(^3P)$<br>$2[O(^3P) + O_2 + M \rightarrow O_3 + M]$<br>$2[O_3 + h\nu \rightarrow O_2 + O(^1D)]^{\#}$<br>$2[H_2O + O(^1D) \rightarrow OH + OH]$<br>$2[H_2 + OH \rightarrow H_2O + H]$<br>$2[H + O_2 + M \rightarrow HO_2 + M]$<br>$\underline{2[OH + HO_2 \rightarrow H_2O + O_2]}$<br>net: $O_2 + 2H_2 \rightarrow 2H_2O$ | 11.0% | Similar to pathway 2 but with $O(^1D)$ removed by $H_2O$ instead of $H_2$ |
| Pathway 4<br>$O_2 + h\nu \rightarrow O(^3P) + O(^3P)$<br>$2[O(^3P) + O_2 + M \rightarrow O_3 + M]$<br>$2[O_3 + H \rightarrow OH + O_2]^{\#}$<br>$\underline{2[H_2 + OH \rightarrow H_2O + H]}$<br>net: $O_2 + 2H_2 \rightarrow 2H_2O$ | 4.2% | Similar to pathway 2 except $O_3$ reacts with H instead of photolysing |
| Pathway 5<br>$CH_4+OH\rightarrow CH_3+H_2O^{\#}$<br>$CH_3+O_2+M\rightarrow CH_3O_2+M$<br>$CH_3O_2+OH\rightarrow H_3CO+HO_2$<br>$H_3CO+O_2\rightarrow H_2CO+HO_2$<br>$H_2CO+OH\rightarrow H_2O+HCO$<br>$HCO+O_2\rightarrow HO_2+CO$<br>$CO+OH\rightarrow CO_2+H$<br>$H+O_2+M\rightarrow HO_2+M$<br>$4[HO_2+O(^3P)\rightarrow OH+O_2]$<br><u>$2[O_2+hv\rightarrow O(^3P)+O(^3P)]$</u><br>net: $2O_2+CH_4\rightarrow 2H_2O+CO_2$ | 2.9% | Oxidation of $CH_4$ by $O_2$ into $H_2O$ and $CO_2$ catalysed by HOx; pathway does not involve $H_2$ |
| Pathway 6<br>$O_2+hv\rightarrow O(^3P)+O(^3P)$<br>$2[NO_2+O(^3P)\rightarrow NO+O_2]$<br>$2[NO+HO_2\rightarrow NO_2+OH]^{\#}$<br>$2[H_2+OH\rightarrow H_2O+H]$<br><u>$2[H+O_2+M\rightarrow HO_2+M]$</u><br>net: $O_2+2H_2\rightarrow 2H_2O$ | 1.8 | Oxidation of $H_2$ into $H_2O$ catalysed by NOx and HOx |
| Pathway 7<br>$O_3+hv\rightarrow O_2+O(^1D)$<br>$O(^1D)+N_2\rightarrow O(^3P)+N_2$<br>$HO_2+O(^3P)\rightarrow OH+O_2$<br>$H_2+OH\rightarrow H_2O+H^{\#}$<br><u>$H+O_2+M\rightarrow HO_2+M$</u><br>net: $O_3+H_2\rightarrow O_2+H_2O$ | 1.5% | Oxidation of $H_2$ into $H_2O$ by $O_3$ catalysed by HOx |

Table 4: Pathway Analysis output for the atmospheric model at ~14km i.e. where gas-phase oxidation of $H_2$ by $O_2$ is most efficient for run 2 (SE x1000$H_2$ run) (see Figure 7). *shown as a %of the total Oy loss rate over the interval analysed. "M" indicates any third-body gas-phase species required to carry away excess vibrational energy of the reactants. [#] indicates the slowest (bottleneck) reaction in the sequence.

Table 4 suggests that the main gas-phase removal of $O_2$ is via conversion into $H_2O$, a process which is catalysed by HOx (pathways 1-4) or by mixed HOx-NOx cycles e.g. (pathway 6). Note that pathway 2 differs from the others in that the $H_2$ is broken by $O^1D$ instead of OH. A smaller (2.9%) contribution (pathway 5) arises from reduction of $O_2$ by $CH_4$. Pathway 7 is particular, in the sense that it is overall a sink for Oy since it converts three atoms of oxygen ($O_3$) into two atoms ($O_2$) (plus one atom of oxygen in water which is not included in the Oy definition). Figure 9 summarises the pathways shown in Table 4 for photochemical gas-phase (H₂-O₂) oxidation:

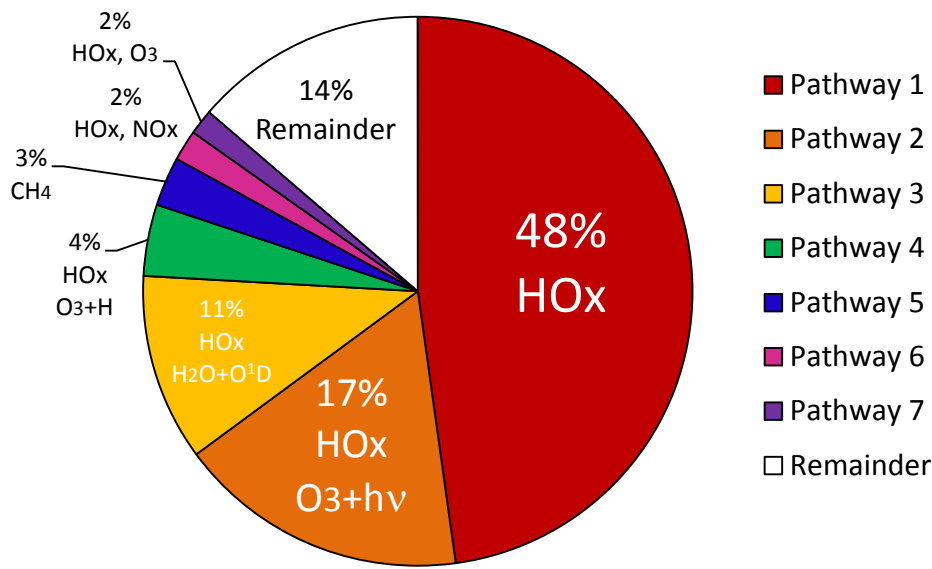

Figure 9: Pie chart summarising the %contribution to the atmospheric Oy photochemical removal rate at ~14km (see Table 4) for the seven pathways found by the pathway analysis program for scenario 2 (SE x1000 H₂ run).

### 6.5 Timescales for O₂ Abiotic Production and Photochemical Removal

We calculate here the timescale for abiotic oxygen production (in section 6.5.1) and the timescale for photochemical removal of O₂(g) by H₂(g) (in section 6.5.2) for run 2 (SE x1000 H₂ run). Comparing these two timescales (in section 6.5.3) indicates whether O₂(g) could build-up to reach the CE limit or whether it is quickly removed by photochemical [H₂-O₂] oxidation.

**6.5.1 Abiotic O$_2$ Production Timescale**

We assume the mechanism of Luger and Barnes (2015) who proposed an abiotic production rate of up to 25 bar O$_2$(g)/Myr for Earth-like planets orbiting M-dwarfs during their Pre-main Sequence Phase. Assuming mass of Earth's atmosphere (NASA Earth factsheet, nssdc.gsfc.nasa.gov):

$$M_{atm\_earth}=5.1x10^{18}kg=5.1x10^9Tg$$

Next, we calculate the mass of molecular oxygen in Earth's atmosphere which is the product of the mass mixing ratio (mmr) of oxygen multiplied by the total atmospheric mass:

Mass O$_2$(g) Earth's atmosphere: $M_{atm\_o2\_earth}=[mmr_{o2}]* M_{atm\_earth}=[0.21*(32/28.8)]*5.1x10^9=1.19x10^9Tg$

The total mass O$_2$(g) in the 3g (P$_o$=1bar)  SE atmosphere (run 2) equals one third the mass of the Earth (1g) case because the higher SE gravity leads to collapse of the atmospheric column at constant surface pressure. This means:

$$M_{atm\_o2\_SE}=(1/3)*M_{atm\_o2\_earth}=3.97x10^8Tg$$

The desired rate of abiotic oxygen production (25 bar oxygen from Luger and Barnes, 2015) is assumed to equal:

$$R_{abiotic} \sim 25*M_{atm\_o2\_SE}=(25/0.21)*3.97x10^8Tg/Myr = 4.72x10^{10}Tg/Myr$$

[We assume thereby that the abiotic rate of "25 bar O$_2$/Myr" as quoted in Luger and Barnes (2015) corresponds to x25 times the SE (run 2) O$_2$ atmospheric mass/Myr].

Finally The corresponding timescale for abiotic  O$_2$(g) production:

$$\tau_{O2\_abio} \sim M_{atm\_o2\_SE} / R_{abiotic} = (3.97x10^8 \, Tg) / (4.72x10^{10} \, Tg/Myr)$$

$$\sim 8400 \text{ years}$$

**6.5.2 Photochemical Oy removal by $H_2(g)$ Timescale in Run 1 and Run 2**

Here we calculate Oy removal times for the relatively modest conditions in run 2 (3g SE with x1000 $H_2$ otherwise modern Earth conditions). The net photochemical removal of Oy by $H_2$ in the region of interest (10-20km, see Table A2 and Figure 8) is $-4.22 \times 10^{12}$ molecules $cm^{-2}$ $s^{-1}$. This is equivalent to a photochemical Oy removal rate:

$R_{Oy\_loss} = -1.45 \times 10^{11}$ Tg/yr assuming a 3g SE with two Earth radii.

Therefore, $\tau_{Oy\_loss} \sim M_{atm\_o2\_SE} / R_{Oy\_loss} = (3.97 \times 10^8 \text{ Tg}) / (1.45 \times 10^1 \text{ Tg/Myr})$

~ 2740 years

6.5.3 Comparison of $O_2(g)$ Production and Loss Timescales

The above analysis suggests that abiotic $O_2$ production has a lifetime ($\tau_{O2\_abio}$) of ~8400 years for the extreme conditions in 6.5.1. Our model results for the more modest conditions (run 2) suggest photochemical Oy removal timescales via $H_2$ ($\tau_{Oy\_H2}$) of ~2740 years, 6.5.2]. This suggests that $H_2$ oxidation has relatively rapid photochemical timescales which can prevent abiotic build-up of $O_2$ in the SE atmospheres considered. An important caveat however, is that our model is only valid for modest $H_2$ amounts i.e. we assume vmr $H_2(g)=5.5 \times 10^{-4}$ in run 2 (x1000 modern Earth) in order to remain within the validity range. A new model version currently being developed for $H_2$-dominated atmospheres to study the more extreme conditions during the pre-main sequence will be the focus of future work.

This result has important potential repercussions. First, for the interpretation of $O_2$ as a biosignature since our work suggests that an important, proposed abiotic source of $O_2$ (Luger and Barnes, 2015) would be strongly weakened in SE atmospheres which have more than a few % of $H_2$. Second, our analysis suggests that CE in such atmospheres could be limited due to a lack of $O_2$. Note however, our work represents a straightforward, global mean approach. Also, abiotic $O_2$ production could be enhanced by $CO_2$ photolysis – a process not considered in our timescale analysis. These issues should be the subject of future work with models valid over a wider compensation and [T,p] range.

**7.0 (CO-$O_2$) Mixtures**

In addition to ($H_2$-$O_2$) there is a wide range of systems which can potentially combust, including mixtures of carbon-containing species in oxygen. In this section we consider the combustion limits of one such system, namely (CO-$O_2$). CO is a key species determining the carbon budget. Its ratio to $CH_4$ is well-studied and helps constrain (C/O) hence the evolution of the star-planet system. In Earth's atmosphere, important sources of CO include biomass burning and in-situ oxidation of hydrocarbons (Pétron et al., 2004). On Mars and Venus CO is produced mainly photochemically (see e.g. Lellouch et al., 1991). In this section we consider the potential of explosion-combustion to affect ($CO-O_2$) abundances in exoplanetary atmospheres of Earth-like and Mini Gas Planets.

## 7.1 $CO-O_2$ Combustion Limits

CO combustion in $O_2$ has been proposed (e.g. Cohen, 1992) although the detailed mechanism is generally not as well understood as for $H_2-O_2$ mixtures. The overall (net) reaction is:

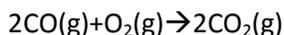

$$2CO(g)+O_2(g)\rightarrow 2CO_2(g)$$

CO combusts in air for abundances between about (16-70%) at room temperature and between about (12-74%) at 300$^o$C (Cohen, 1992, their Figure 10 and references therein).  In damp atmospheres, it is likely that HOx resulting from $H_2O$ photolysis would catalyze CO into $CO_2$ so the CO is less likely to build up to its combustive limit.

## 7.2 Application to Earth-like and Mini Gas Planets (MGPs)

On modern Earth, CO atmospheric abundances at the surface vary from ~(30-120) ppbv depending on latitude and season (Khalil and Rasmussen, 1994). For an Earth-like planet in the HZ of an M-dwarf star, this value could rise by (2-3) orders of magnitude (Segura et al., 2005) but still lies far below the CE limit. The rise in CO is due to a slowing in the reaction: $CO+OH\rightarrow CO_2+H$ due to low OH. The low OH arises because the reaction: $O(^1D)+H_2O\rightarrow 2$ OH is weak, since $O(^1D)$ production from ozone photolysis is weak due to weak UV emission in the relevant wavelength range from the central star.

Regarding MGPs, the model study of Hu and Seager (2014) (their Figures 5 and 6) varied e.g. C/O ratios and predicted atmospheric compositions which suggested MGPs could form with atmospheric concentrations of several tens of percent by volume of CO and $O_2$. Their results were averaged from p=(1000-100)mb and T from about (700-800)K. Inspecting the CE limit for CO (see Figure 10 and the discussion below) suggests that these atmospheres would combust, although due to the large parameter range (in terms of e.g. metallicity, central star etc.) more studies are needed to investigate the full range of effects.  In their Figure 5 for a GJ1214b like planet, the combustion limit for atmospheric ($CO-O_2$) is reached – with the CO vmr exceeding ~10% and the $O_2$ vmr  reaching up to 20% - for C/O values ranging from (0.3-0.5) and for $X_H$ ranging from (0.2-0.5) (see the panels in their Figure 5 marked CO and $O_2$). In their Figure 6 for a 55 Cnc e-like planet the combustion limit for ($CO-O_2$) is similarly reached  – again with CO and $O_2$ vmrs of up to 20% -for C/O values ranging from (0.2-0.6) and for $X_H$ ranging from (0.0-0.6). The study by Miguel and Kaltenegger (2014) although mainly focusing on hot mini-Neptunes also provided model results with and without disequilibrium (photochemistry and mixing) processes (see e.g. their Figure 8) for cooler (down to 700K), hydrogen-dominated atmospheres with C/O=0.54. Their work suggested that photochemistry becomes important in the upper atmosphere regions at pressures less than ~0.1 bar. At greater pressures, $CO(g)$ forms thermochemically and the influence of photochemistry is negligible. This suggests an important difference between ($CO$-$O_2$) combustion and ($H_2$-$O_2$) combustion: whereas $CO(g)$ is produced by equilibrium chemistry at pressures greater than ~ 0.1bar, abiotic $O_2(g)$ however, is likely produced either via photochemistry at pressures smaller than ~ 0.1bar or possible thermochemically at greater pressures under certain conditions(see discussion on Hu and Seager study above). Thermochemically-produced $CO(g)$ at such pressures is not affected by photochemical removal e.g. via HOx-catalysed oxidation into $CO_2(g)$.

## 8.0 Hydrocarbon-$O_2$-$N_2$ Mixtures

Hydrocarbons (e.g. $CH_4$) can constitute an important part of the atmospheric carbon budget especially for planets which orbit beyond the ice-line where colder atmospheric temperatures mean that reduced forms of carbon are thermodynamically favored. In this section we investigate the potential of CE to affect the abundances of hydrocarbon-$O_2$-$N_2$ mixtures. The lower and upper limits of CE for different gases, namely $H_2$, CO, $CH_4$, ethylene ($C_2H_4$) and propane ($C_3H_8$) with air as a fill gas were determined by Zlochower and Green (2009) (see their Table 1) as summarized below in Figure 10:

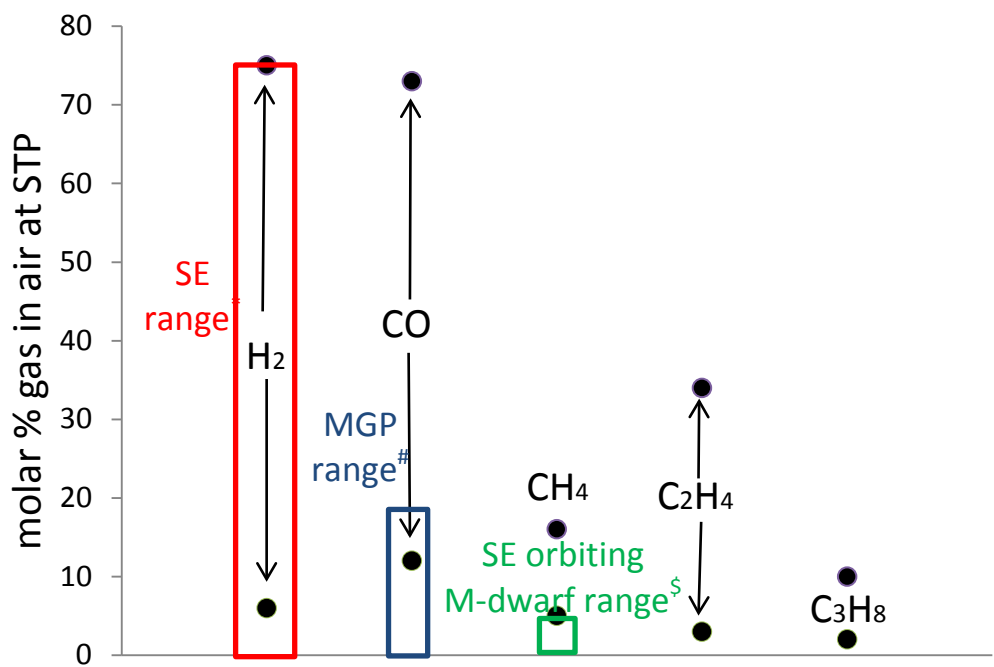

Figure 10: Combustion-explosion range shown by the black arrows for the molar concentration of five gases determined in air under Standard Temperature and Pressure (STP) conditions by Zlochower and Green (2009). Red, blue and green rectangles show the range of possible atmosphere compositions for SEs, MGPs and SEs orbiting in the HZ of M-dwarf stars respectively. *See Figure 1 and accompanying text. #See section 7. $See section 8.

(CH$_4$(g)-O$_2$(g)) mixtures can combust-explode with net products depending on the relative amounts of reacting gases as follows:

CH$_4$ (g)+O$_2$(g)→CO$_2$(g) +2H$_2$(g) (low oxygen)

2CH$_4$ (g)+3O$_2$(g)→2CO(g) +4H$_2$O(g) (medium oxygen)

CH$_4$ (g)+2O$_2$(g)→CO$_2$(g) +2H$_2$O(g) (excess oxygen)

Figure 10 suggests that CH$_4$(g) undergoes CE in air at STP for molar concentrations ranging from 4% by mole (the "lean limit") up to 16% by mole (the "rich limit"). This corresponds to a lower limit for O$_2$(g) of 6% by mole. At higher temperatures the lower (lean) limit decreases by 0.4% by mole for each 100K increase in temperature (Gieras et al., 2006). This suggests that a typical warm SE (with T=700K) would have a $CH_4(g)$ lean limit of 2.4% by mole (vmr) at one bar.

What $CH_4(g)$ vmr concentrations are predicted in the Earth-like atmospheric literature? Model studies of such planets orbiting in the HZ of cool stars (Segura et al., 2005; Rauer et al., 2011) which assume Earth's biomass predict enhanced $CH_4(g)$ concentrations compared with modern Earth - but still not enough by at least an order of magnitude for $(O_2(g)-CH_4(g))$ explosion-combustion to occur. In these scenarios, lowered UV output from the star weakens photolytic hydroxyl radical $(OH(g))$ production which is the main sink for $CH_4(g)$. Grenfell et al. (2014) varied biomass emissions and incoming stellar UV for an Earth-like planet orbiting in the mid HZ of cool M-dwarf stars and calculated one scenario - for a quiet, cool M7 star which featured 2.7% $CH_4(g)$ by vmr – which may combust, if the atmosphere were much warmer. That study also calculated four further scenarios where $CH_4(g)$ by vmr exceeded ~0.5%. Rugheimer et al (2015) studied even cooler (up to M9) M-dwarf cases but held the surface $CH_4(g)$ in their model constant. For their (M6-M9) spectral cases this approach was equivalent to assuming rather weak surface $CH_4(g)$ biomass emissions of ~x100 times weaker than on Earth. In summary, only a few scenarios in the literature so far predict that the $(O_2(g)-CH_4(g))$ CE could be approached - for SE atmospheres orbiting stars with spectral class M7 and cooler. Nevertheless, the full parameter range is not explored. Also, Earth's biomass is in some studies reduced in order to remain within the model's validity range. The outer HZ range for Earth-like planets orbiting cooler stars, a region where low UV is expected to favour $CH_4(g)$ build-up is not well explored.

A caveat when simulating atmospheres with abundant $CH_4(g)$ is that organic aerosols can start to form when the $CH_4(g)$ vmr exceed a few tenths of a percent depending on temperature and $CO_2(g)$ (see e.g. Trainer et al., 2004; Zerkle et al., 2012). Regarding higher volatile organic compounds (e.g. C1-C3) – these species combust in oxygen at threshold abundances which are about x5 times lower than methane (see e.g. Figure 10; see also Gas Data Book, 2001). More studies are required to investigate this issue further.

### 9.0 $H_2$-$CH_4$-$NH_3$-$N_2O$-$O_2$-$N_2$ Mixtures

We briefly note here that atmospheric species which are found on Earth and on gas giants - such as ammonia ($NH_3$) as well as the Earth biosignature nitrous oxide ($N_2O$) – could both undergo combustion reactions in mixtures of $H_2$-$CH_4$-$NH_3$-$N_2O$-$O_2$-$N_2$ (Pfahl et al., 2000) although the details of the chemical and physical mechanism are not well known. The molar concentrations required for combustion (at least a few %) for these two species are however likely not reached in most currently-conceivable exoplanetary atmospheric scenarios since e.g. $NH_3$ sources are weak and since this molecule is removed via e.g. photolysis and rainout quite quickly (typically on the order of hours to days on modern Earth). Also for $N_2O$ the atmospheric sources hence the molar concentrations are usually rather low (~$3 \times 10^{-7}$ on modern Earth).

N$_2$O is also a product for CO in air and CH$_4$ in air combustion in ($O_2$-$N_2$) mixtures. Malte and Pratt (1974) (see also Steele et al., 1995) reported formation of several ppmv $N_2O(g)$ e.g. via the reaction $N_2+O+M \rightarrow N_2O+M$ for gas mixtures near the lean limit for CO-air combustion from (0.5-1.0) bar. The yield of $N_2O$ depends on the combustion temperature since $N_2O$ is thermally-decomposed. The study by Park et al. (2004) suggested formation via the reaction: $NH+NO \rightarrow N_2O+H$. Summarizing, the issue of $N_2O$ formation by combustion requires further work but has potentially important repercussions for interpreting $N_2O$ as an exoplanetary biosignature.

## 10. Dust explosions

Suspended dust can present a large surface area of combustible material which can lead to atmospheric explosions at much lower threshold values in gas mixtures than would occur without the presence of dust. The dust explosion threshold is sensitive to particle size (typically <100micron diameters are required) and needs a minimum dust loading which typically varies between (10-50) g/m$^3$ for many organic materials. For more information refer to Amyotte (2013). We mention this phenomenon only briefly here for the sake of completeness. In the context of SE atmospheres however, data on dust or aerosol amounts etc. are not available - although first clues of the possible presence of strong aerosol loadings are one possible interpretation for the rather featureless atmospheric spectra of some mini gas planets.

## 11. Discussion and Conclusions

CE could in certain cases constrain the range of atmospheric compositions in exoplanetary atmospheres although photochemical oxidation of $H_2$ by $O_2$ likely plays an important role in limiting the build-up of $O_2$. To investigate these initial findings further, more work is required to examine responses over the potentially wide range of composition, p, and T using consistent (1D and 3D) models which investigate cases where the CE limit could be reached considering gas-phase chemistry, escape etc. Our initial analysis suggests that the accumulation of abiotic $O_2$ as proposed by Luger and Barnes (2015) could be prevented due to CE or/and photochemical oxidation of $H_2$ by $O_2$. This has important repercussions for interpreting $O_2(g)$ as a biosignature although further studies are needed. Future work includes the development of a coupled climate-photochemical model which can simulate conditions approaching the limit (~a few percent by volume mixing ratio depending on (p,T) where $H_2$-$O_2$ combustion-explosion take place. This will require updating the radiative transfer in the climate module as well as expanding the $H_2$ photochemistry reaction network and modifying H- and O-fluxes at the model upper boundary.

The pathway analysis suggested that photochemical oxidation of [$H_2$-$O_2$] operates mainly in a relatively narrow altitude range in the middle atmosphere – high enough such that HOx (and NOx) are released from their reservoirs but low enough such that the product $H_2O(g)$ is not photolysed. CE could provide a means of re-distributing the atmospheric energy budget by converting chemical energy in the atmosphere into other energy forms (e.g. heat, radiance, sound) which could favor more rapid atmospheric cooling hence the formation of planetary oceans.

$CO(g)$-$O_2(g)$ mixtures could potentially reach the combustion-explosion threshold for a sub-set of mini gas planets and SEs with the appropriate metallicity in the T range (600-800)K for p>1 bar where thermochemical production of CO dominates. An important caveat is that $O_2(g)$ only builds-up thermochemically for $X_H$ <0.5 since hydrogen otherwise reduces $O_2(g)$. For SEs having $X_H$>0.5 therefore, abiotic $O_2(g)$ production would likely proceed mainly via photochemistry. Whether significant CO can form photochemically e.g. via photolytic release from $CO_2(g)$ requires further studies investigating timecales of e.g. HOx-catalysed photochemical regeneration of $CO(g)$ into $CO_2(g)$ which depends on the UV environment and the atmospheric moisture content. This was investigated for Mars by e.g. Stock et al. (2012).

($CH_4$-$O_2$) mixtures in the current literature e.g. considering planets with Earth's biomass and development moved to the HZ of (F, G,K, M) main sequence stars, $CH_4(g)$ remains below the limit for CE. For the M-dwarf star cases (e.g. M0-M5), $CH_4(g)$ builds-up to more than x1000 that on modern Earth - but this is still about a factor of five below the explosion-combustion limit at 1bar. Nevertheless, there are still important scenarios which are not yet explored, where much higher $CH_4(g)$ abundances are expected – possibly exceeding the combustion limit. These include Earth-like planets in the mid to outer HZ (where UV is low which favors the build-up of $CH_4(g)$) and for such planets orbiting the coolest (M7-M9) M-dwarf stars. In the literature, such scenarios apply only very low $CH_4(g)$ emissions (~1% of the modern Earth). Initial tests (not shown) with our coupled photochemical-climate model for Earth-like planets orbiting in the HZ of M-dwarf stars where we explored the mid to outer HZ and also the effect of varying $CH_4(g)$ biomass emissions in the range (1-10) times the modern Earth, suggested that the $CH_4(g)$ concentration rapidly approached the combustion limit for low UV conditions. Further work however is needed to extend our models to be valid for higher $CH_4$(g) abundances before investigating this issue further.

**Appendix 1: Effect of radical removal via sticking on solid surfaces**

The limits of CE are frequently determined in the laboratory using reaction chambers. Sticking collisions (hence removal) of reactive of gas-phase radicals on the inner walls of the chamber disfavor CE. To estimate the role of surface chemistry for planetary atmospheres, Table A1 shows the ratio surface area divided by the volume of gas (atmosphere) for a range of conditions:

| Earth's troposphere[*] | Reaction chamber[#] | Stratospheric aerosol[$] | Polluted troposphere[&] | Martian global dust devil[##] |
|---|---|---|---|---|
| $9.98 \times 10^{-5}$ | 3.00 | $1.00 \times 10^{-7}$ | $1.00 \times 10^{-4}$ | $4.83 \times 10^{-4}$ |

Table A1: The ratio surface area divided by the volume of gas (atmosphere) for a range of conditions. [*]Value represents the volume shell from Earth's surface up to z=10km altitude divided by the total surface area (ocean plus continents) of the Earth assuming a spherical planet. [#]Assuming a spherical chamber with 2m diameter. [$]Value represents the mean stratospheric sulfate aerosol loading of the modern Earth (Seinfeld and Pandis, 2006). [&]Schryer, 1982. [##]Assuming 1000 dust particles cm$^{-3}$ with a radius of 1.6 microns (Esposito et al., 2011).

Values in Table A1 suggest that atmospheric scenarios feature lower surface/volume ratios than reaction chambers used in the laboratory to determine the conditions for CE. The first and third explosion limits can be sensitive to surface reactions (Wang and Law, 2013) – in atmospheres the rather low surface areas in Table A1 suggest that these limits would therefore be reached more easily (at lower p, T) in planetary atmospheres compared with the laboratory determined limits. Experimental data is however lacking so further quantification of the conditions where the first and third limits would be reached is the focus of future work.

## Appendix 2: Total Oxygen (Oy) Removal Rate Profiles

Table A2 shows the Total Oxygen (Oy), (= $[2O_2+3O_3+O(^3P)+O(^1D)+OH+2HO_2+2H_2O_2+2ClO_2+ClO+NO+2NO_2]$) (see section 6.3) removal rate (ppbv/s) in run 2 arising due to gas-phase pathways which oxidize $H_2(g)$ into $H_2O(g)$ (see Figure 5 and Table 2) in the middle atmosphere:

| Model layer[#] | Density (molecules/cm$^3$) | $\Delta$Oy(ppbv/s) |
|---|---|---|
| 10 | $9.68 \times 10^{18}$ | 0.0 |
| 12 | $7.24 \times 10^{18}$ | 0.0 |
| 14 | $5.36 \times 10^{18}$ | -0.0001 |
| 16 | $3.93 \times 10^{18}$ | -0.0002 |
| 18 | $2.87 \times 10^{18}$ | -0.0003 |
| 20 | $2.10 \times 10^{18}$ | -0.001 |
| 22 | $1.54 \times 10^{18}$ | -0.002 |
| 24 | $1.13 \times 10^{18}$ | -0.003 |
| 26 | $8.28 \times 10^{17}$ | -0.005 |
| 28 | $6.09 \times 10^{17}$ | -0.007 |
| 30 | $4.50 \times 10^{17}$ | -0.011 |
| 32 | $3.34 \times 10^{17}$ | -0.015 |
| 34 | $2.49 \times 10^{17}$ | -0.024 |
| 36 | $1.85 \times 10^{17}$ | -0.033 |
| 38 | $1.37 \times 10^{17}$ | -0.046 |
| 40 | $1.03 \times 10^{17}$ | -0.052 |
| 42 | $7.79 \times 10^{16}$ | -0.050 |
| 44 | $5.94 \times 10^{16}$ | -0.045 |
| 46 | $4.54 \times 10^{16}$ | -0.036 |
| 48 | $3.47 \times 10^{16}$ | -0.028 |
| 50 | $2.66 \times 10^{16}$ | -0.022 |
| 52 | $2.02 \times 10^{16}$ | -0.016 |
| 54 | $1.53 \times 10^{16}$ | -0.012 |
| 56 | $1.15 \times 10^{16}$ | -0.009 |
| 58 | $8.56 \times 10^{15}$ | -0.007 |
| 60 | $6.40 \times 10^{15}$ | -0.004 |

Table A2: Oy (= $[2O_2+3O_3+O(^3P)+O(^1D)+OH+2HO_2+2H_2O_2+2ClO_2+ClO+NO+2NO_2]$) removal rate (ppbv/s) due to gas-phase pathways for run 2 in the atmospheric column model. [#]Model layers extend 10 (~3km) up to layer 60 (~20km).